\documentclass[conference]{IEEEtran}
\IEEEoverridecommandlockouts
\def\BibTeX{{\rm B\kern-.05em{\sc i\kern-.025em b}\kern-.08em T\kern-.1667em\lower.7ex\hbox{E}\kern-.125emX}}

\usepackage{graphicx} 
\usepackage{amsmath}
\usepackage{caption}
\usepackage{amsfonts}
\usepackage{bm}
\usepackage{enumitem}
\usepackage{acronym}
\usepackage{cite}

\usepackage{multirow}
\usepackage{microtype}
\usepackage{balance}
\usepackage{glossaries}
\usepackage[utf8]{inputenc}
\usepackage{multirow}
\usepackage{hhline}
\usepackage{colortbl}
\usepackage{multicol}
\usepackage{diagbox}  
\usepackage{booktabs} 
\usepackage{makecell} 
\usepackage{graphicx}    
\usepackage[font=footnotesize,labelformat=simple]{subcaption} 
\usepackage{comment} 

\usepackage[a-1b]{pdfx}

\usepackage[font = footnotesize, textfont=up]{caption}
\newacronym{cpp}{CPP}{carrier phase positioning}
\newacronym{cnn}{CNN}{convolutional neural network}

\newacronym{bie}{BIE}{best integer equivariant}
\newacronym{tof}{ToF}{time-of-flight}
\newacronym{psd}{PSD}{power spectral density}
\newacronym{csi}{CSI}{channel state information}
\newacronym{nn}{NN}{neural network}
\newacronym{los}{LoS}{line-of-sight}
\newacronym{cf}{CF}{cell-free}
\newacronym{ml}{ML}{machine learning}
\newacronym{ue}{UE}{user equipment}
\newacronym{mimo}{MIMO}{multiple-input multiple-output}
\newacronym{mle}{MLE}{maximum likelihood estimation}
\newacronym{dl}{DL}{deep learning}
\newacronym{ap}{AP}{antenna point}
\newacronym{flop}{FLOP}{floating-point operation}
\newacronym{toa}{ToA}{time-of-arrival}
\newacronym{mlp}{MLP}{multi-layer perceptron }
\newacronym{rmse}{RMSE}{Root mean-squared error}
\newacronym{relu}{ReLU}{Rectified Linear Unit}
\newacronym{conv2d}{Conv2D}{2D-Convolutional}
\newacronym{mse}{MSE}{mean-squared error}
\newacronym{ecdf}{ECDF}{empirical cumulative distribution function}
\newacronym{srs}{SRS}{sounding reference signal}

\def\BibTeX{{\rm B\kern-.05em{\sc i\kern-.025em b}\kern-.08em
    T\kern-.1667em\lower.7ex\hbox{E}\kern-.125emX}}
\begin{document}
\bstctlcite{IEEEexample:BSTcontrol}
\title{Phase-Only Positioning: Overcoming Integer Ambiguity Challenge through Deep Learning
}

\author{\IEEEauthorblockN{Fatih Ayten\IEEEauthorrefmark{1}, Mehmet C. Ilter\IEEEauthorrefmark{1}, Ossi Kaltiokallio\IEEEauthorrefmark{1}, Jukka Talvitie\IEEEauthorrefmark{1}, Akshay Jain\IEEEauthorrefmark{2},\\ Elena Simona Lohan\IEEEauthorrefmark{1}, Henk Wymeersch\IEEEauthorrefmark{3}, and Mikko Valkama\IEEEauthorrefmark{1} }  
\IEEEauthorblockA{
\IEEEauthorrefmark{1} Electrical Engineering Unit, Tampere Wireless Research Center, Tampere University, Finland\\
\IEEEauthorrefmark{2} Radio Systems Research, Nokia Bell Labs, Espoo, Finland\\
\IEEEauthorrefmark{3}Department of Electrical Engineering, Chalmers University of Technology, Sweden\\
}}

\maketitle

\begin{abstract}
This paper investigates uplink \gls{cpp} in \gls{cf} or distributed antenna system context, assuming a challenging case where only phase measurements are utilized as observations.
In general, \gls{cpp} can achieve sub-meter to centimeter-level accuracy but is challenged by the integer ambiguity problem. In this work, we propose two deep learning approaches for phase-only positioning, overcoming the integer ambiguity challenge. The first one directly uses phase measurements, while the second one first estimates integer ambiguities and then integrates them with phase measurements for improved accuracy. Our numerical results demonstrate that an inference complexity reduction of two to three orders of magnitude is achieved, compared to maximum likelihood baseline solution, depending on the approach and parameter configuration. This emphasizes the potential of the developed deep learning solutions for efficient and precise positioning in future \gls{cf} 6G systems.

\end{abstract}


\glsresetall    

\section{Introduction}
%
\Gls{cpp} is an established approach in global navigation satellite systems \cite{10536135,ding2021carrier}, and is currently receiving growing interest also in mobile networks towards centimeter-level terrestrial positioning \cite{Cha2025, 3gpp_tr_38_859_2024,10475845,5g_carrier_phase_prs_1}.
%
\gls{cpp} is subject to the so-called integer ambiguity challenge, as the mapping of the transmitter-receiver distance to the observable carrier phase is invariant to any integer amounts of wavelengths \cite{10536135,10232971}.
Existing solutions include differential phase measurements, hybrid time-phase estimation, and multipath-assisted ambiguity resolution \cite{8955972}. Additionally, while traditional \gls{mle} methods suffer from high computational complexity \cite{726808}, data-driven techniques leveraging \glspl{nn} and clustering-based estimators for ambiguity resolution have been recently developed \cite{gnss_ambiguity_est_1, gnss_ambiguity_est_2}. 

In terrestrial networks, \gls{cpp} is relevant in the context of distributed \gls{mimo} and \gls{cf} systems -- an important and emerging paradigm towards 6G  \cite{10536135,cell_free_massive_mimo,10379539}. \gls{cf} and other distributed \gls{mimo} systems eliminate traditional cell boundaries by distributing large numbers of transmission/reception or \glspl{ap}
over wide areas. 
Especially with phase-coherent \glspl{ap}, such a distributed approach offers tempting prospects also for sensing and localization in the effective near-field domain \cite{10851402,s19204582}.

\textls[-4]{In general, most existing \gls{cpp} studies such as \cite{10437902, Cha2025, 10475845,5g_carrier_phase_prs_1} combine carrier phase measurements as \emph{additional observations}, on top of more common positioning-related measurements like \gls{toa}. The same applies also implicitly to the direct positioning methods described in \cite{10851402,s19204582,10694279} which harness directly the raw I/Q received signals for localization purposes -- embedding the complete \gls{csi}. To the best of the authors' knowledge, there are no existing terrestrial radio positioning works that rely exclusively on carrier phase measurements for localizing the \gls{ue}. The only exception is  near-field \gls{ue} localization using phase-difference type of measurements, however, assuming a large co-located antenna array with classical half-wavelength element spacing, such as \cite{9508850}.}


\begin{figure}[t!] 
    \centering
    \vspace{-0.5mm}
    \includegraphics[width=0.4\textwidth]{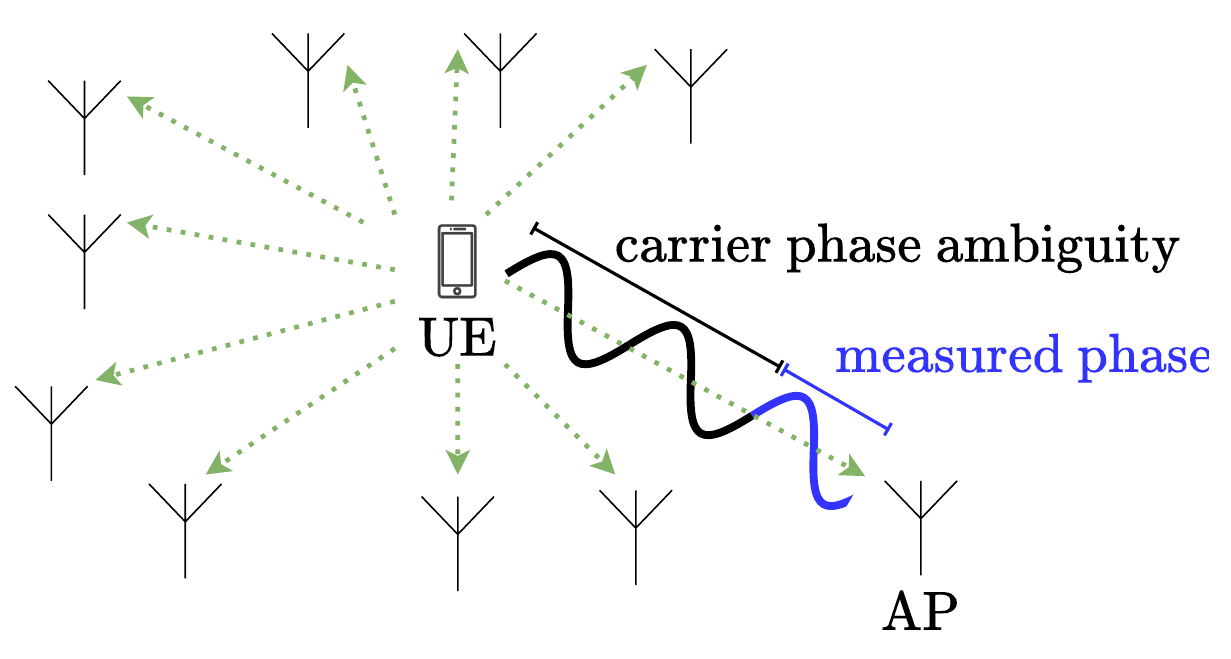}
    \caption{Uplink positioning with distributed antenna points (APs) where only the carrier phase measurements at different APs are used to estimate the UE position.}\vspace{-0mm}
    \label{fig:system_model}
\end{figure}

Inspired by the above, this paper seeks to fill this important gap and addresses the \emph{phase-only \gls{ue} positioning paradigm} and the related integer ambiguity challenge in \gls{cf}/distributed \gls{mimo} system context -- illustrated conceptually in Fig.~\ref{fig:system_model}. Utilizing only the carrier phase measurements offers remarkable implementation benefits, as \gls{toa} and other related \gls{csi} measurements are always subject to hardware impairments such as the inevitable \gls{ue} clock bias \cite{10536135,10437902}.
To this end, two alternative deep learning based \gls{ue} positioning approaches are proposed harnessing only the distributed carrier phase measurements.
The first approach, called direct phase-only \gls{cpp}, utilizes a \gls{mlp} to directly infer position estimates from raw carrier phase measurements. The second approach, entitled integer ambiguity-aided \gls{cpp}, employs a \gls{cnn} with a dedicated \gls{mlp}-based module for integer ambiguity estimation, which is then combined with phase data for improved localization accuracy. To assess their practical feasibility, we analyze the computational complexity of both approaches in terms of \glspl{flop} and compare them with more ordinary \gls{mle}-based reference approach. Our results demonstrate that the proposed deep learning approaches not only outperform traditional \gls{mle}-based methods in terms of the achievable positioning accuracy for fixed processing complexity in terms of \glspl{flop}, but also obtains really high centimeter-level positioning accuracy with largely improved 
computational efficiency. These findings pave the way for real-time carrier phase-based positioning in future 6G distributed MIMO or cell-free networks.

\section{System Model and Baseline}
We consider an uplink scenario consisting of a \gls{ue} and $I$ distributed and mutually phase-synchronized \glspl{ap}. An example scenario with ten \glspl{ap} is shown in Fig. \ref{fig:system_model}. The \gls{ue} has an unknown position $\boldsymbol{x}_{\text{ue}} \in \mathbb{R}^2$ whereas the \glspl{ap} have known positions $\boldsymbol{x}_{\text{ap,}i} \in \mathbb{R}^2$ for $i \in \{0, \dots I-1\}$, with the scenario being extendable to $\mathbb{R}^3$.  When the \gls{ue} transmits a unit-power narrowband pilot symbol $s$ occupying a bandwidth $W$ with a transmit power $P$, the received signal at the $i$-th \gls{ap} over the \gls{los} link is $ y_i = \sqrt{E} \rho_i \exp({-j(\frac{2\pi}{\lambda}d_i-\theta)})s+v_i$, 
where $E=P/W$ is the symbol energy, $\rho_i$ is the path loss from the \gls{ue} to \gls{ap} $i$, 
$d_i =  \Vert\boldsymbol{x}_{\text{ue}}-\boldsymbol{x}_{\text{ap,}i}\Vert$ is the Euclidean distance, $\theta$ is the common phase offset between the \gls{ue} and the \gls{ap} network, and $v_i \sim \mathcal{CN}(0,N_0)$ represents additive white complex Gaussian noise. By processing $y_i$, the resulting phase observations 
can be formulated as
\begin{equation}\label{eq:phase_estimations}
    r_i = -\frac{2\pi}{\lambda}d_i+\theta+ 2 \pi z_i  + n_i,
\end{equation}
where $z_i \in \mathbb{Z}$ is the integer ambiguity, and we model $n_i \sim \mathcal{N}(0,\sigma_i^2)$. From \eqref{eq:phase_estimations}, we can take one \gls{ap} as reference, e.g.,  $i = 0$, and compute \emph{differential measurements} as
\begin{equation}\label{eq:differential_measurements}
    \delta_m = -\frac{\lambda}{2\pi}(r_m-r_0) = \Delta_m+k_m\lambda+w_m, m \in \{1, \dots ,I-1\},
\end{equation}
where $\Delta_m=d_m-d_0$, $k_m=z_0-z_m$ and $w_m=(\lambda/2\pi)(n_0-n_m)$. It is also assumed that the maximum delay difference between the reference \gls{ap} and the other \glspl{ap} is much smaller than $1/W$. Based on Fisher information theory, the lower bound on the error covariance of $\boldsymbol{w} = [w_1, \ldots w_{I-1}]^\top$ can now be shown to read
\begin{equation} \label{eq:noise_covariance_lower_bound}
   \mathbf{\Sigma}_\text{diff}= \frac{\lambda^2 N_0}{8\pi^2E} (\mathbf{D}+\frac{\mathbf{1}\mathbf{1}^\top}{\rho_0^2}),
\end{equation}
where $\mathbf{D}$ is a diagonal matrix whose $m$-th diagonal element is given by $[\mathbf{D}]_{m,m}=1/{\rho_m^2}$, $\mathbf{1}$ is a vector of $I-1$  ones and the superscript $(.)^\top$ denotes the transpose operation. Due to space constraints, the derivation of 
\eqref{eq:noise_covariance_lower_bound} is omitted.

As a baseline, we will devise and  use the \gls{mle}. 
The vector of differential measurements in \eqref{eq:differential_measurements} can be expressed as
\begin{equation}\label{eq:differential_measurements_vector}
\boldsymbol{\delta} = \boldsymbol{h}(\boldsymbol{x}_{\text{ue}}) + \boldsymbol{w},
\end{equation}
where $\boldsymbol{\delta}\!=\!\left[\delta_1,\!\ldots\!\delta_{I-1}\right]^\top$, $\boldsymbol{h}(\boldsymbol{x}_{\text{ue}})\!=\!\left[h_1(\boldsymbol{x}_{\text{ue}}), \! \ldots h_{I-1}(\boldsymbol{x}_{\text{ue}})\right]^\top$, $h_m(\boldsymbol{x}_{\text{ue}}) = \frac{-\lambda}{2\pi}(\text{mod} (-2\pi \Vert\boldsymbol{x}_{\text{ue}}-\boldsymbol{x}_{\text{ap,}m}\Vert/\lambda,2\pi)  -  \text{mod}(-2\pi \Vert\boldsymbol{x}_{\text{ue}}-\boldsymbol{x}_{\text{ap,}0}\Vert/\lambda,2\pi))$ and $\text{mod}(a, b)$ denotes the modulo operation of $a$ with respect to $b$.
Given that $\boldsymbol{w}$ follows a multivariate normal distribution with zero-mean and covariance matrix $\boldsymbol{\Sigma}_\text{diff}$, the likelihood function is $\mathcal{L}(\boldsymbol{\delta} | \boldsymbol{x}_{\text{ue}}) = \mathcal{N}(\boldsymbol{\delta}; h(\boldsymbol{x}_{\text{ue}}), \boldsymbol{\Sigma}_\text{diff})$. The \gls{mle} for the \gls{ue} position can be obtained by minimizing the negative log-likelihood, given by
\begin{equation}\label{eq:final_mle_equation}
\begin{aligned}
\widehat{\boldsymbol{x}}_{\text{ue}}= \underset{\boldsymbol{x}_{\text{ue}}}{\arg \min}  \ & 
(\boldsymbol{\delta} -\boldsymbol{h}(\boldsymbol{x}_{\text{ue}}))^\top \boldsymbol{\Sigma}_\text{diff}^{-1} (\boldsymbol{\delta} -  \boldsymbol{h}(\boldsymbol{x}_{\text{ue}})).
\end{aligned}
\end{equation}
The \gls{mle} has a notable computing complexity, quantified in Section IV, while serving as the performance benchmark.

\section{Proposed \gls{nn}-based Positioning Approaches} \label{section:nn_models}
In this section, we introduce two \gls{nn}-based positioning approaches, utilizing a total of three \glspl{nn}. The first approach (referred to as direct phase-only CPP) employs a single \gls{mlp} model that directly processes raw phase measurements to estimate the \gls{ue} position. The second approach (referred to as integer ambiguity-aided CPP), consists of two \glspl{nn}: an \gls{mlp}-based model that estimates the integer ambiguities from phase measurements, and a \gls{cnn} model that integrates these estimated ambiguities with phase measurements to determine the \gls{ue} position.

\subsection{Direct Phase-Only Carrier Phase Positioning}\label{subsection:carrier_phase_positioning}
We first propose a straightforward approach to estimate the \gls{ue} position using a \textit{\gls{mlp}-based positioning model}. The proposed \gls{mlp} takes the differential measurements $\boldsymbol{\delta}$ in \eqref{eq:differential_measurements_vector} as input and generates an estimate $\widehat{\boldsymbol{x}}_{\text{ue}}$ of the true \gls{ue} position $\boldsymbol{x}_{\text{ue}}$. The loss function employed in this model is the \gls{mse}, expressed as $\Vert\boldsymbol{x}_{\text{ue}}-\widehat{\boldsymbol{x}}_{\text{ue}}\Vert^2$.
The proposed model is a fully connected \gls{nn} composed of dense layers. The model consists of an input layer with a dimension of $I-1$ and \gls{relu} activation, followed by seven hidden layers with dimensions of $A$, $2A$, $4A$, $8A$, $4A$, $2A$, and $A$, all using \gls{relu} activation, and an output layer with a dimension of 2 and linear activation function.

\subsection{Integer Ambiguity-Aided Carrier Phase Positioning}
The second approach first estimates the integer ambiguities using a \gls{mlp}-based ambiguity estimation model. The estimated ambiguities are then combined with phase measurements and fed into a \gls{cnn}-based positioning model.

\subsubsection*{MLP-based ambiguity estimation model}\label{subsubsection:mlp_based_ambiguity_estimation_model}
We propose a \gls{mlp}-based \gls{nn} to generate a set of probability distributions $\boldsymbol{P} = \left[ \boldsymbol{p}_1 , \ldots \boldsymbol{p}_{I-1} \right]^\top$ for the differential ambiguities $\boldsymbol{k} = \left[k_1, \ldots, k_{I-1} \right]^\top$ from the differential measurements $\boldsymbol{\delta}$.
Each differential ambiguity $k_m$ is geometrically bounded by $\left[-q_m,q_m \right] = \left[ -\lfloor \Vert \boldsymbol{x}_{\text{ap,}m}-\boldsymbol{x}_{\text{ap,}0}\Vert/\lambda \rfloor , \lfloor \Vert\boldsymbol{x}_{\text{ap,}m}-\boldsymbol{x}_{\text{ap,}0}\Vert/\lambda \rfloor \right]$, where $\lfloor \cdot \rfloor$ denotes the floor function that returns the largest integer less than or equal to its argument. This results in $Q_m = 2 q_m +1$ possible labels for each differential ambiguity, with the total number of possible labels across all differential ambiguities given by $Q=\sum_{m=1}^{I-1} Q_m$.
For each $k_m$, the model generates a probability distribution 
$\boldsymbol{p}_m = \left[ p_{m,-q_m} , ... p_{m,q_m} \right]^\top$, 
where $p_{m,l}\in [0,1]$ represents the probability of the differential ambiguity $k_m$ 
taking the integer value $l$, with $\sum_l p_{m,l}=1$.

\begin{figure}
    \centering
    \vspace{-0em}
    \includegraphics[width=0.45\textwidth]{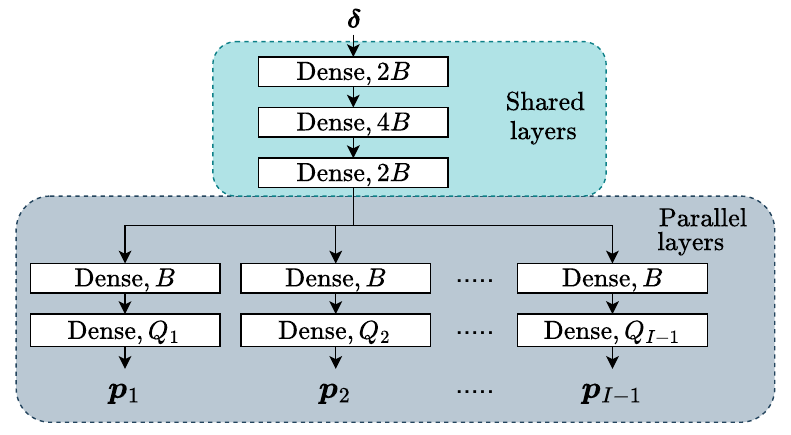}
    \caption{Proposed \gls{mlp}-based ambiguity estimation model: Rectangles represent dense layers with the number of neurons inside each rectangle and each branch generates a probability distribution.}\vspace{-1em}
    \label{fig:ambiguity_estimator_figure}
\end{figure}

The proposed \gls{mlp} structure is illustrated in Fig.~\ref{fig:ambiguity_estimator_figure}. The input first propagates through shared layers, then the model is divided into parallel branches. The shared layers and the first layer of the parallel branches utilize the \gls{relu} activation function, whereas the output layers of the parallel branches utilize the softmax activation function to output the probability distributions.

The sparse categorical cross-entropy is employed as the loss function, which can be expressed as
\begin{equation}
\mathcal{L}_{\text{SCCE}} = -\frac{1}{I-1} \sum_{m=1}^{I-1} \ln(p_{m,k_m}),
\end{equation}
where $p_{m,k_m}$ is the predicted probability of the correct integer label for each differential ambiguity and $\ln(.)$ is the natural logarithm.

\subsubsection*{CNN-based positioning model}\label{subsubsection:cnn_based_positioning_model}
We introduce a positioning model based on a \gls{cnn} that estimates the \gls{ue} position by integrating $\boldsymbol{\delta}$ with the estimated ambiguities $\boldsymbol{\widehat{k}} = [\widehat{k}_1, \ldots, \widehat{k}_{I-1}]^\top$. These estimated ambiguities are obtained from the \gls{mlp}-based ambiguity estimation model described in Section \ref{subsubsection:mlp_based_ambiguity_estimation_model}, where the output probability distributions are processed through an argmax operation to yield the final integer estimates. Specifically, the argmax operation determines the predicted integer label for each differential ambiguity branch by selecting the class with the highest probability, i.e., $\widehat{k}_m = \arg\max_{l \in [-q_m, q_m]} p_{m,l}$.
 
\begin{figure}
    \centering
    \includegraphics[width=0.47\textwidth]{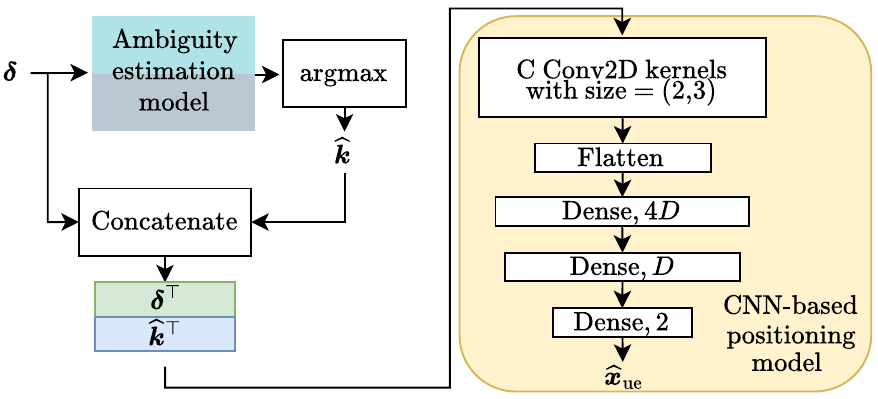}
    \caption{Proposed \gls{cnn}-based positioning model incorporating an \gls{mlp}-based ambiguity estimation model and a variety of layer types.}\vspace{-0em}
    \label{fig:cnn_model_figure}
\end{figure}

Fig.~\ref{fig:cnn_model_figure} illustrates the proposed \gls{cnn} model structure. By noting that both $\boldsymbol{\delta}$ and $\widehat{\boldsymbol{k}}$ vectors have the shape of $(I-1,1)$, the concatenated matrix $[\boldsymbol{\delta}, \widehat{\boldsymbol{k}}]^\top$ has the shape of $(2,I-1)$. The \gls{conv2d} layer with \gls{relu} activation applies $C$ different learnable filters with a shape of $(2,3)$ to the input data, using zero padding and a stride of 1. 
The output of the \gls{conv2d} layer has the shape of $(1,I-3,C)$, which is then flattened to a vector of shape $(C(I-3),1)$. Subsequently, dense layers with \gls{relu} activations are applied. Finally, the output layer with linear activation produces the \gls{ue} location estimate. Similar to the \gls{mlp}-based positioning model, \gls{mse} serves as the loss function.

\section{Inference Complexity}
In this section, the inference complexities of the \gls{mle}-based and \gls{nn}-based approaches are evaluated. The metric used for this purpose is the \gls{flop} count, which is a widely adopted metric \cite{pruning3,flop_motivation1,flop_motivation2,flop_motivation3,flop_motivation4,flop_count_reference,flop_count_reference_2}. In our analysis, each operation—whether an addition, a subtraction, or a multiplication—is counted as one \gls{flop} following the approach in \cite{flop_count_reference,flop_count_reference_2}.

Network pruning, the task of reducing the size of a network by selectively removing parameters, has gained significant attention to tackle over-parameterization and redundancy in deep learning models \cite{pruning1,pruning2,pruning3}. In this work, layerwise magnitude-based pruning \cite{pruning4} is applied, where a portion of weight parameters with the lowest absolute value are pruned in each layer. This approach ensures that the sparsity level remains consistent across all layers. Moreover, if the \glspl{flop} associated with bias additions and activation function in \gls{nn} approaches are neglected, the pruning ratio can be directly applied to the number of \glspl{flop} for each proposed \gls{nn}. Note that, since the \gls{cnn}-based positioning model incorporates the \gls{mlp}-based ambiguity estimation model, the pruning ratios of both models should be explicitly considered when calculating the number of \glspl{flop} in the \gls{cnn}-based positioning model.

\subsection{Maximum Likelihood Estimation Complexity}
The task of finding $\widehat{\boldsymbol{x}}_{\text{ue}}$ that maximizes the \gls{mle} function can be accomplished via a 2D grid search over the simulation area. For simplicity, we neglect the complexity associated with calculating the term $\boldsymbol{h}(\boldsymbol{x}_{\text{ue}})$ in \eqref{eq:final_mle_equation} and assume that $\boldsymbol{\Sigma}_\text{diff}^{-1}$ is precomputed, focusing solely on the computational complexity of the term $(\boldsymbol{\delta} - \boldsymbol{h}(\boldsymbol{x}_{\text{ue}}))^\top \boldsymbol{\Sigma}_\text{diff}^{-1} (\boldsymbol{\delta} - \boldsymbol{h}(\boldsymbol{x}_{\text{ue}}))$.
This operation entails $(I-1)$ subtractions, $I(I-1)$ multiplications, and $I(I-2)$ additions, resulting in a total computational cost of $2I^2 - 2I - 1$ \glspl{flop}. When a 2D grid search is performed over $N_{\text{grid}}$ points, the total computational cost for the \gls{mle} becomes
\begin{equation}\label{eq:mle_flop}
   \mathcal{C}_{\text{MLE}}= N_{\text{grid}}(2I^2 - 2I - 1).
\end{equation}

\subsection{MLP-based Positioning Model Complexity}
A dense layer with $n_i$ input features and $n_o$ neurons requires approximately $n_o(2n_i - 1) \approx 2n_o n_i$ \glspl{flop}, with bias addition and activation costs being negligible for layers containing many neurons. As explained in Section \ref{subsection:carrier_phase_positioning}, the \gls{mlp}-based positioning model is composed entirely of dense layers. Therefore, the number of \glspl{flop} with a pruning rate of $\rho_{\text{MLP}}$ is approximately $ \mathcal{C}_{\text{MLP}}\approx \rho_{\text{MLP}} 168A^2$.

\subsection{MLP-based Ambiguity Estimation Model Complexity}
Similar to the \gls{mlp}-based positioning model, the \gls{mlp}-based ambiguity estimation model described in Section \ref{subsubsection:mlp_based_ambiguity_estimation_model} also consists of only dense layers. The total number of \glspl{flop} with a pruning rate of $\rho_{\text{AE}}$ can be calculated and expressed as
\begin{equation}\label{eq:ambiguity_estimation_model_flops}
    \mathcal{C}_{\text{AE}} \approx \rho_{\text{AE}} \left[32B^2 + (4B^2\!+\!4B) \times (I\!-\!1) + 2B \times Q \right].
\end{equation}

\vspace{-0.5mm}
\subsection{CNN-based Positioning Model Complexity} \label{sec:cnn_model_complexity}
Applying an input data with shape $(2, I-1)$ to $C$ filters each with shape $(2, 3)$ in the \gls{cnn}-based positioning model in Section \ref{subsubsection:cnn_based_positioning_model} costs $\mathcal{C}_{\text{CNN,c}} = 11C \times (I-3)$ \glspl{flop}, where a dot product between a $(2, 3)$ filter and the corresponding input window costs $11$ \glspl{flop}. The number of \glspl{flop} of the subsequent dense layers can be calculated as $\mathcal{C}_{\text{CNN,d}}  \approx 8D^2+4D+8CD(I-3)$.
Also, since the \gls{cnn}-based positioning model requires $\widehat{\boldsymbol{k}}$ to function as shown in Fig. \ref{fig:cnn_model_figure}, the complexity expression of the ambiguity estimation model in \eqref{eq:ambiguity_estimation_model_flops} should also be considered. With a pruning rate of $\rho_{\text{CNN}}$ in the \gls{conv2d} layer and the subsequent dense layers, the total number of \glspl{flop} is given by
\begin{equation}
    \mathcal{C}_{\text{CNN}} = \rho_{\text{CNN}}\times(\mathcal{C}_{\text{CNN,c}} + \mathcal{C}_{\text{CNN,d}}) + \mathcal{C}_{\text{AE}}.
\end{equation}

\vspace{0.0mm}
\section{Numerical Results}
We consider a square-shaped evaluation area of \( 100 \, \text{m}^2 \). A total of $I=20$ antenna points 
are uniformly random distributed over the area in order to stimulate a \gls{cf} network architecture \cite{cell_free_massive_mimo}, while two alternative deployment frequencies of 800\,MHz and 1.8\,GHz are considered. Similary, one \gls{ue} is placed uniformly at random in the area. In our simulation scenario, the total number of possible labels across all differential ambiguities, $Q$, is calculated as 660 for a center frequency of 800\,MHz and 1472 for 1.8\,GHz. The layer parameters of the \gls{nn} models are set as $A = B = D = 128$ for dense layers and $C = 32$ for convolutional filters.

\textls[-1]{The other evaluation parameters include noise \gls{psd} of --174\,dBm/Hz, receiver noise figure of 13\,dB, and uplink transmit powers of $\{-30, -20, -10, 0\}$\,dBm. Also, the free space path-loss model is employed. The uplink pilot waveforms follow the 5G-NR \gls{srs} specifications outlined in \cite{3gpp_TS_38211}. The center frequency is set to either 800\,MHz or 1.8\,GHz, with a subcarrier spacing of 15\,kHz. The pilot transmission utilizes 4 resource blocks and a comb factor of 4, resulting in a utilized uplink reference signal bandwidth of 180 kHz. A single reference symbol configuration \cite{3gpp_TS_38211} is used, and the modulation scheme employed is BPSK.}

\vspace{-0.0mm}
\subsection{Training of NN Models}
The proposed \glspl{nn} are trained separately for each transmit power and center frequency configuration, using $700 \times 10^3$ training samples and $150 \times 10^3$ validation samples per configuration. At each sample, a randomly drawn \gls{ue} location is used. As a standard way of training procedure in supervised learning, true labels are used during the training phase of models, trained according to the input-output relationship of models explained in Section \ref{section:nn_models}.

The models are trained using a batch size of 1000 for 1000 epochs. The Adam optimizer is employed with a learning rate of $10^{-4}$, and L2 regularization is applied with a coefficient of $10^{-5}$. 
To decrease inference complexity, pruning is applied during training. The training process begins with an initial phase of 100 epochs without pruning. After this phase, between epochs 100 and 400, a polynomial decay pruning scheduler is employed, gradually pruning $\rho\%$ of the weight parameters, starting from an initial sparsity of zero and reaching a final sparsity of $\rho\%$ \cite{pruning5}. After pruning, the sparse network is retrained until epoch 1000, allowing the network to adjust and compensate for the removed weight parameters. Each model is first trained without pruning. Then, in subsequent training trials, the pruning rate is increased by steps of $25\%$, with the pruning rate being increased to the point where model performance remains largely unaffected. The optimal pruning rate varies based on the complexity of the problem for each frequency and \gls{nn} model. 

\textls[-1]{Table~\ref{table:pruning_rates} presents the pruning rates and the corresponding \gls{flop} counts for the two proposed approaches, along with the \gls{nn} models included at the two frequencies. As discussed in Section \ref{sec:cnn_model_complexity}, for a given frequency, the \gls{flop} count of the \gls{cnn}-based positioning model includes the \gls{flop} count of the \gls{mlp}-based ambiguity estimation model at the same frequency.}

\begin{table}[!t]
\centering
\caption{\textsc{Pruning Rates and Inference FLOP numbers}}
\label{table:pruning_rates}
\begin{tabular}{|c|c|c|c|}
\hline
\multicolumn{4}{|c|}{\textbf{Direct Carrier Phase-Based Positioning Approach}} \\
\hline
NN Model & Frequency & \begin{tabular}[c]{@{}c@{}}Pruning\\Rate\end{tabular} & \begin{tabular}[c]{@{}c@{}}FLOP\\number\end{tabular} \\
\hline
MLP-based positioning & 800 MHz & 50\% & $1.376 \times 10^6$ \\
\hline
MLP-based positioning & 1.8 GHz & 50\% & $1.376 \times 10^6$ \\
\hline
\hline
\multicolumn{4}{|c|}{\textbf{Integer Ambiguity-Aided Carrier Phase Positioning Approach}} \\
\hline
NN Model & Frequency & \begin{tabular}[c]{@{}c@{}}Pruning\\Rate\end{tabular} & \begin{tabular}[c]{@{}c@{}}FLOP\\number\end{tabular} \\
\hline

\begin{tabular}[c]{@{}c@{}}MLP-based ambiguity\\estimation\end{tabular} & 800 MHz & 50\% & $0.974 \times 10^6$ \\
\hline
CNN-based positioning & 800 MHz & 75\% & $1.148 \times 10^6$ \\
\hline
\begin{tabular}[c]{@{}c@{}}MLP-based ambiguity\\estimation\end{tabular} & 1.8 GHz & 0\% & $2.156 \times 10^6$ \\
\hline
CNN-based positioning & 1.8 GHz & 75\% & $2.330 \times 10^6$ \\
\hline
\end{tabular}
\end{table}

\vspace{-0.5mm}
\subsection{Parametrization of MLE Benchmarks}
As a first approach, to ensure a fair comparison with the proposed \gls{nn} approaches, the number of grid points (\(N_{\text{grid}}\) in \eqref{eq:mle_flop}) should be chosen to match the \gls{flop} number shown in Table~\ref{table:pruning_rates}. After finding the required $N_{\text{grid}}$ values, they are rounded up to ensure that $N_{\text{grid}}$ is a perfect square, allowing the grid points to be distributed uniformly in a square area with equal spacing in both dimensions.

As an alternative approach, we calculate the complexity reduction factor of the proposed \glspl{nn} by determining the required $N_\text{grid}$ values needed to achieve comparable positioning performance. 
In both MLE approaches, after identifying the grid point that maximizes the \gls{mle}, a gradient descent algorithm with 100 steps is used to fine-tune the estimation.

\vspace{-0.5mm}
\subsection{Ambiguity Estimation Results}
To evaluate the performance of the \gls{mlp}-based ambiguity estimation model, we define two accuracy metrics, namely element-wise accuracy and overall accuracy, over $T$ test samples. The element-wise accuracy is given by
\begin{equation}
    \text{Acc}_e = \frac{1}{T} \frac{1}{I-1} \sum_{t=1}^{T}  \sum_{m=1}^{I-1} \mathbb{I}(\widehat{k}_{m,t} = k_{m,t} )  \times 100\%,
\end{equation}
where $\widehat{k}_{m,t}$ and $k_{m,t}$ are the estimated and true ambiguities $\widehat{k}_{m}$ and $k_{m}$ for the test sample $t$, respectively, and $\mathbb{I}(\cdot)$ is the indicator function, which equals 1 when its argument is true and 0 otherwise.
The overall accuracy is, in turn, defined as
\begin{equation}
    \text{Acc}_o = \frac{1}{T} \sum_{t=1}^{T} \mathbb{I}(\widehat{\boldsymbol{k}}_t = \boldsymbol{k}_t)  \times 100\%,
\end{equation}
where $\widehat{\boldsymbol{k}}_t$ and $\boldsymbol{k}_t$ are the estimated and true ambiguity vectors for the test sample $t$, respectively.

\textls[-1]{The accuracy results, evaluated over $T = 150 \times 10^3$ test samples for the considered frequencies and UE transmit power levels, are presented in Table~\ref{tab:ambiguity_estimation_performance}. The accuracy decreases at the higher frequency of 1.8\,GHz compared to 800\,MHz, which can be attributed to the increased number of integer ambiguities that must be resolved. However, overall, the obtained accuracies are high, reflecting accurate ambiguity resolution.}

\begin{table}[t]
\centering
\caption{\textsc{Accuracy Performance of Ambiguity Estimator Models}}
\label{tab:ambiguity_estimation_performance}
\begin{tabular}{l l *{4}{c}}
\toprule
\multirow{2}{*}{} & & \multicolumn{4}{c}{\textbf{Transmit Power [dBm]}} \\  
\cmidrule{3-6}
\textbf{Metric} & \textbf{Frequency} & --30 & --20 & --10 & 0 \\
\midrule
\multirow{2}{*}{$\text{Acc}_e$} & 800\,MHz & 98.97\% & 99.57\% & 99.83\% & 99.89\% \\
& 1.8\,GHz & 92.64\% & 97.90\% & 99.08\% & 99.43\% \\
\midrule
\multirow{2}{*}{$\text{Acc}_o$} & 800\,MHz & 83.72\% & 92.97\% & 97.24\% & 98.40\% \\
& 1.8\,GHz & 32.74\% & 71.13\% & 86.48\% & 91.59\% \\
\bottomrule
\end{tabular}
\end{table}

\subsection{Positioning Results}
\begin{figure}[t!]
\centering
\captionsetup[subfigure]{labelformat=empty}
\begin{subfigure}[b]{0.42\textwidth}
    \centering
   \includegraphics[width=1\linewidth]{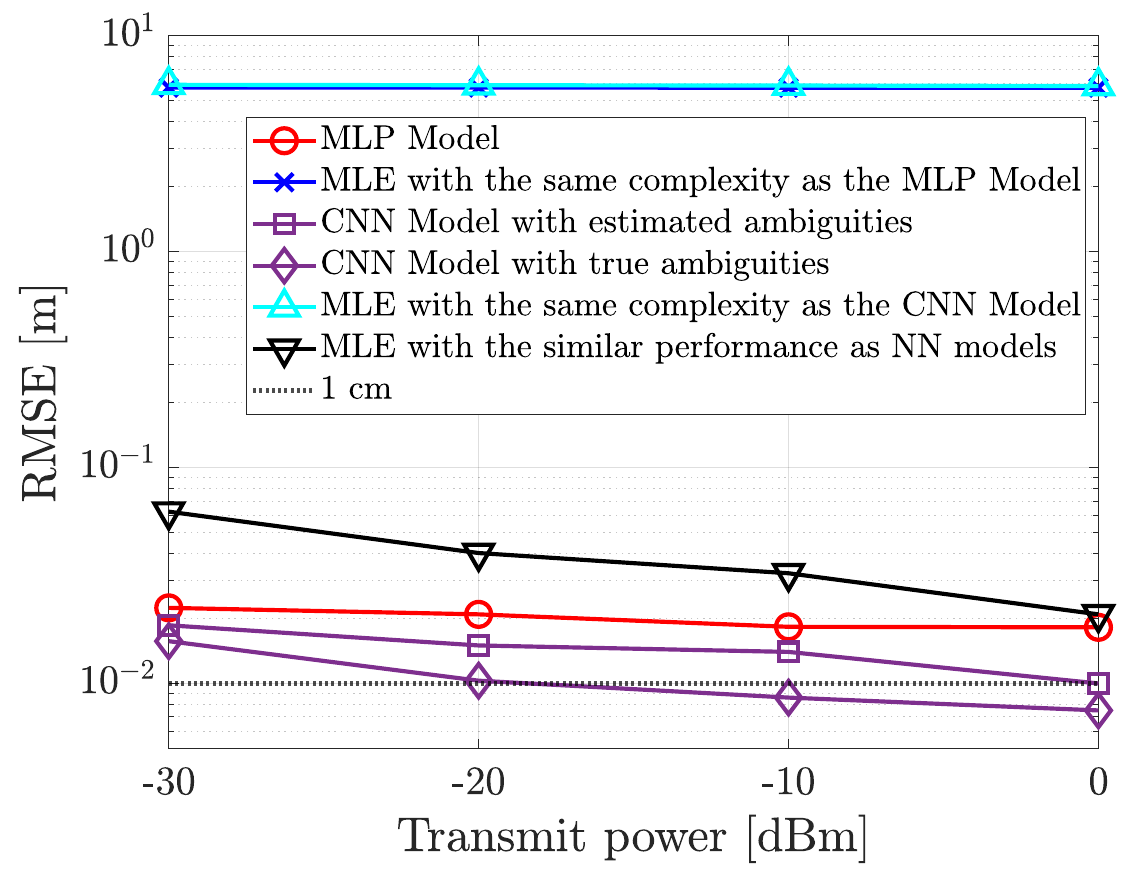}
   \vspace{-5mm}
   \subcaption{(a) 800\,MHz distributed deployment}
   \label{fig:positoning_error_800_MHz} 
\end{subfigure}
\vspace{-1em}
\begin{subfigure}[b]{0.42\textwidth}
    \centering
   \includegraphics[width=1\linewidth]{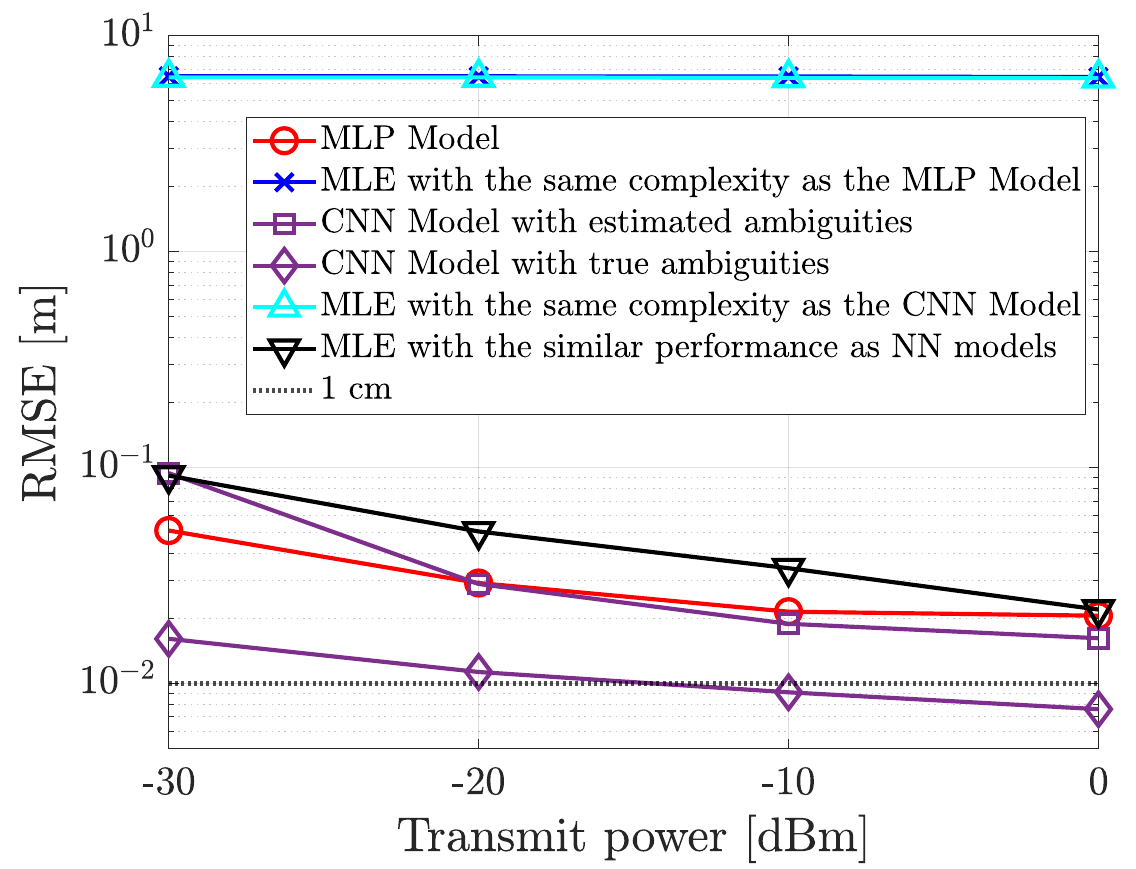}
   \vspace{-5mm}
   \subcaption{(b) 1.8\,GHz distributed deployment}
   \label{fig:positoning_error_1.8_GHz}
\end{subfigure}
\vspace{2mm}
\caption{Positioning accuracy results and comparison between the two proposed \glspl{nn} and their \gls{mle} counterparts.}
\label{fig:positoning_error}\vspace{-1mm}
\end{figure}

\textls[-10]{\gls{rmse}-based positioning results are calculated based on $150 \times 10^3$ test samples. A comparison of the proposed \gls{mlp}-based and \gls{cnn}-based positioning models with the corresponding \gls{mle} counterparts, is presented in Fig. \ref{fig:positoning_error} for frequencies of 800\,MHz and 1.8\,GHz.
The impact of ambiguity estimation on \gls{cnn}-based positioning model performance was evaluated by comparing scenarios utilizing true ambiguities versus estimated ambiguities. Both \gls{cnn}- and \gls{mlp}-based models demonstrate superior performance compared to their respective \gls{mle} counterparts, each matched in inference complexity. This superiority can be attributed to \gls{mle}'s limitations under constrained grid search parameters designed to maintain comparable inference complexity across \gls{nn} models.}

As illustrated in Fig. \ref{fig:positoning_error}, the \gls{cnn}-based model outperforms the \gls{mlp}-based model despite reduced inference complexity as seen in Table~\ref{table:pruning_rates} at 800\,MHz. Notably, the \gls{cnn}-based model employing estimated ambiguities achieves a \gls{rmse} of 1\,cm at 0\,dBm transmit power. Fig.~\ref{fig:positoning_error} reveals that while the \gls{cnn}-based model exhibits lower performance than the \gls{mlp}-based model at low transmit power, it outperforms the \gls{mlp}-based model when transmit power exceeds --20\,dBm at 1.8\,GHz. However, at 1.8\,GHz, the \gls{cnn}-based model requires greater inference complexity compared to the \gls{mlp}-based model (cf. Table~\ref{table:pruning_rates}). Additionally, the performance difference between \gls{cnn} models utilizing estimated versus true ambiguities is more pronounced at 1.8\,GHz, primarily due to the somewhat reduced accuracy of the ambiguity estimation model at this frequency (cf. Table~\ref{tab:ambiguity_estimation_performance}).

\textls[-1]{The required $N_\text{grid}$ values for the MLE to achieve similar performance as the proposed \glspl{nn} are found to be $750^2 = 0.5625 \times 10^6$ and $1800^2 = 3.24 \times 10^6$ for 800\,MHz and 1.8\,GHz, respectively.  
For \gls{mlp}-based positioning, the inference complexity reduction factors are approximately 310 at 800\,MHz and 1787 at 1.8\,GHz, while for \gls{cnn}-based positioning, the factors are around 372 and 1055 at 800\,MHz and 1.8\,GHz, respectively. Due to the need for a very dense grid search at higher frequencies with MLE, the complexity reduction factors through the proposed methods are significantly higher at 1.8 GHz -- being already three orders of magnitude. Furthermore, the optimal \gls{nn} that maximizes the reduction factor varies depending on the frequency.}

\begin{figure}[t!] 
    \centering
    \includegraphics[width=0.4\textwidth]{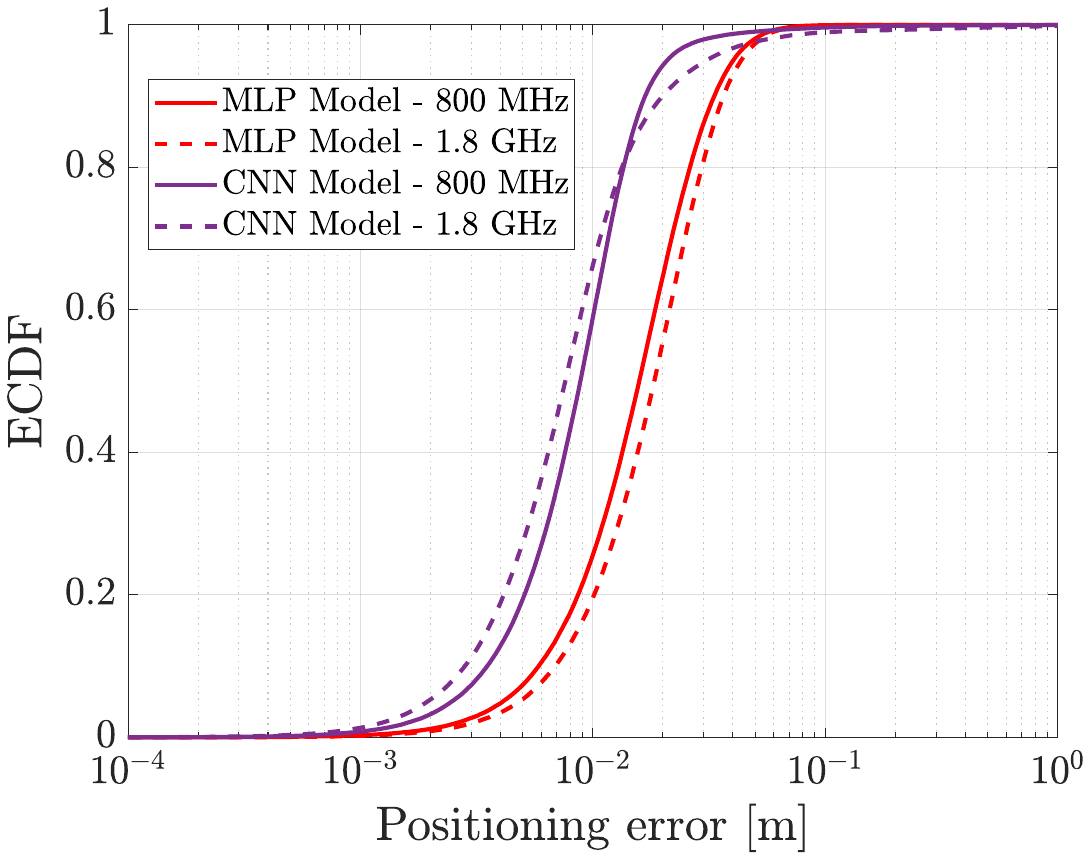}
    \caption{ECDF of the proposed \glspl{nn} at a transmit power of 0\,dBm.}\vspace{-0em}
    \label{fig:ecdf}
\end{figure}

To conduct a more comprehensive analysis, the positioning error \gls{ecdf} curves of the proposed \glspl{nn} at a transmit power of 0\,dBm are illustrated in Fig.~\ref{fig:ecdf}. In the \gls{mlp} model, increasing the frequency notably degrades the positioning performance. In contrast, the \gls{cnn} model demonstrates improved performance at higher frequencies, with more \gls{ecdf} values below approximately 1.4\,cm. Beyond this threshold, the lower frequency exhibits superior performance, which can be attributed to the ambiguity estimation accuracy at these frequencies. The 95th percentile positioning errors for the \gls{cnn} model are 2.08\,cm at 800\,MHz and 3.08\,cm at 1.8\,GHz, whereas the \gls{mlp} model exhibits errors of 4.01\,cm at 800\,MHz and 4.34\,cm at 1.8\,GHz.

\section{Conclusion}
This paper introduced and proposed two deep learning-based approaches for uplink carrier phase positioning that rely exclusively on distributed uplink phase observations in a cell-free system -- allowing to avoid classical \gls{toa} measurements and the related challenges with clock biases. Our results show that both proposed approaches outperform traditional \gls{mle} benchmark method under the constraint of comparable inference complexity. Specifically, the \gls{cnn}-based model achieves superior accuracy at higher transmit power levels and demonstrates a positioning error of approximately 1\,cm at 0\,dBm UE transmit power at 800\,MHz. However, its performance is also more sensitive to ambiguity estimation accuracy at higher frequencies (1.8\,GHz). In contrast, the \gls{mlp}-based model exhibits more stable performance across varying power levels. These findings highlight the potential of deep learning for efficient and precise phase-only positioning in future cell-free systems, paving the way for improved localization in environments where \gls{toa}-based methods are impractical. Our future work will focus on studying and mitigating the impacts of multipath and residual phase calibration errors at the network side.

\balance
\bibliographystyle{IEEEtran}
\bibliography{References}

\end{document}